\documentclass[twocolumn,showpacs,preprintnumbers,floatfix,amsmath,amssymb,superscriptadd
ress]{revtex4}

\usepackage{graphicx}% Include figure files
\usepackage{dcolumn}% Align table columns on decimal point
\usepackage{bm}% bold math

\newcommand{\be}{\begin{equation}}
\newcommand{\ee}{\end{equation}}
\newcommand{\bn}{\begin{eqnarray}}
\newcommand{\en}{\end{eqnarray}}
\newcommand{\ba}{\begin{array}}
\newcommand{\ea}{\end{array}}
\newcommand{\bc}{\begin{center}}
\newcommand{\ec}{\end{center}}
\newcommand{\bml}{\begin{mathletters}}
\newcommand{\eml}{\end{mathletters}}

\begin{document}

\preprint{ }

\title{Continuum Coupling and Spectroscopic Properties of Nuclei} 

\author{N. Michel}
\affiliation{
Department of Physics and Astronomy, University of Tennessee, Knoxville, Tennessee
37996
}%
\affiliation{
Physics Division, Oak Ridge National Laboratory, Oak Ridge, Tennessee 37831
}%
\affiliation{Joint Institute for Heavy Ion Research,
Oak Ridge National Laboratory, P.O. Box 2008, Oak Ridge, Tennessee 37831}

\author{W. Nazarewicz}
\affiliation{
Department of Physics and Astronomy, University of Tennessee, Knoxville, Tennessee
37996
}%
\affiliation{
Physics Division, Oak Ridge National Laboratory, Oak Ridge, Tennessee 37831
}%
\affiliation{
Institute of Theoretical Physics, University of Warsaw, ul. Ho\.za 69,
00-681 Warsaw, Poland }%

\author{M. P{\l}oszajczak}
\affiliation{
Grand Acc\'el\'erateur National d'Ions Lourds (GANIL), CEA/DSM - CNRS/IN2P3,
BP 55027, F-14076 Caen Cedex, France
}%

\author{J. Rotureau}
\affiliation{
Department of Physics and Astronomy, University of Tennessee, Knoxville, Tennessee
37996
}%
\affiliation{
Physics Division, Oak Ridge National Laboratory, Oak Ridge, Tennessee 37831
}%
\affiliation{Joint Institute for Heavy Ion Research,
Oak Ridge National Laboratory, P.O. Box 2008, Oak Ridge, Tennessee 37831}

\date{\today}% It is always \today, today,
             %  but any date may be explicitly specified 

\begin{abstract}
The nucleus is a correlated open quantum many-body system. 
The presence of  states that are unbound to particle emission may have significant 
impact on spectroscopic properties of nuclei.
In the framework of the continuum shell model in
the complex momentum-plane, we discuss salient effects of the
continuum coupling on the spectroscopic factors, one-neutron overlap integrals,
and energies of excited states in $^{6}$He and  $^{18}$O. We demonstrate
the presence of non-perturbative  threshold effects in many-body wave
functions of weakly bound and unbound states.
\end{abstract}


\maketitle

One of the main frontiers of  nuclear science  is the structure of
short-lived,   radioactive nuclei with extreme neutron-to-proton
asymmetry. Such nuclei inhabit outskirts of the chart of the nuclides,
close to the particle drip lines, where the nuclear binding comes to an
end. In this context, the major challenge for nuclear theory is to
develop theories and algorithms that would allow to understand 
properties of those exotic  physical systems 
possessing  new and different properties \cite{doba,brown}. To this end, a unification of structure and
reaction aspects of weakly bound or unbound nuclei,  based on  the open
quantum system (OQS) formalism, is called for.

The nuclear shell model (SM) is the cornerstone of our understanding of
nuclei. In its standard realization \cite{brown,cau}, SM assumes that
the many-nucleon system is perfectly isolated from an external
environment of scattering states and decay channels. The validity of
such a closed quantum system (CQS) framework is sometimes justified by 
relatively high one-particle (neutron or proton) separation energies in
nuclei close to the valley of beta stability. However, weakly bound or
unbound nuclear states cannot be treated in a CQS formalism.  A
consistent description of the interplay between scattering states,
resonances, and bound states in the many-body wave function requires an
OQS formulation (see Refs.~\cite{mah,opr} and references quoted
therein). Properties of unbound states lying above the particle (or
cluster) threshold directly impact the continuum structure. Coupling to
the particle continuum is also important for weakly bound states, such
as halos. A classic example of a threshold effect is the Thomas-Ehrman
shift \cite{te} which manifests itself in the striking asymmetry in the
energy spectra between mirror nuclei. Another example is the so-called
helium anomaly \cite{oglo}, i.e.,  the presence of  higher one- and
two-neutron emission thresholds in $^{8}$He than in $^{6}$He.

In this work, we investigate the impact of the non-resonant continuum on
spectroscopic properties of weakly bound and unbound nuclei. In
particular,  the one-neutron overlap integrals and energies of excited 
states are studied using the SM in the complex $k$-plane, the so-called
Gamow Shell Model (GSM) \cite{Mic02,Mic04,Bet02}. By explicit many-body
calculations that fully account for a   coupling to scattering space, 
we demonstrate the presence of a non-perturbative rearrangement in the
wave function with a significantly low angular momentum single-particle
(s.p.) component \cite{wigner,Op05}.

GSM is the multi-configurational SM with a complex-$k$ s.p. basis given
by the Berggren ensemble \cite{Berggren} consisting of Gamow states
(poles of the s.p. $S$-matrix; sometimes called Siegert or resonant
states) and the  non-resonant continuum of scattering states. For a
given partial wave,$(\ell,j)$, the scattering states are distributed
along the  contour $L_+^{\ell_j}$ in the complex momentum plane. The
Berggren basis is generated by a finite-depth potential, and the
many-body SM states are  the linear combination of Slater determinants
built from the s.p. Berggren  states. The attractive feature of the GSM
is that it  can treat simultaneously both continuum effects and
many-body correlations between nucleons due to the configuration mixing.
All details of the formalism can be found in Ref. \cite{Mic02}.

Single-nucleon overlap integrals and the associated spectroscopic
factors (SFs) are basic ingredients of the theory of direct reactions
(single-nucleon transfer, nucleon knockout, elastic break-up)
\cite{satch,sfs}. Experimentally, SFs  can be  deduced from measured
cross sections; they are useful measures of  the configuration mixing in
the many-body wave function. The associated reaction-theoretical
analysis often reveals model- and probe-dependence \cite{sf1,sf2,sf3}
raising concerns about the accuracy of  experimental determination of
SFs. In our study we discuss the  uncertainty in determining SFs due to 
the two assumptions commonly used in  the standard SM studies, namely
(i) that a nucleon is transferred to/from a specific s.p.  orbit
(corresponding to an observed s.p. state), and  (ii) that the transfer
to/from the continuum of non-resonant scattering states can be
disregarded.

The SF for a  given reaction channel is defined as a real part of 
an overlap between the initial and final states  \cite{sfs}:
\begin{equation}
S^2=\langle [\Psi^{J_{A-1}}_{A-1} \otimes u_{\ell j} ]^{J_A} | 
\Psi^{J_A}_A \rangle^2,
\end{equation} 
where  $| \Psi^{J_{A-1}}_{A-1} \rangle$ and $| \Psi^{J_A}_A \rangle$ are
the  wave functions of nuclei $A$ and $A$$-$1, and  $u_{\ell j}$ is
the wave function  of the transferred nucleon (normalized to unity).
The radial part of $u_{\ell j}$ represents the one-nucleon radial overlap
integral.
 Using a decomposition of the $(\ell,j)$
channel in the complete Berggren basis, one obtains:
\begin{equation}
S^2 =\mathcal{N} \left( 
\int\hspace{-1.4em}\sum_{b} \langle \Psi^{J_{A}}_{A} 
|| a^+_{\ell j}(b) || \Psi^{J_{A-1}}_{A-1} \rangle^2 
\right),
\label{eq3}
\end{equation}
where  $a^+_{\ell j} (b)$ is a
creation operator of a s.p. basis state that is either
a  discrete Gamow state  or a 
scattering continuum state. 
The normalization constant $\mathcal{N}$
is determined  from the condition that $S^2$=1
when $\displaystyle | \Psi^{J_A}_A \rangle = |[\Psi^{J_{A-1}}_{A-1}
\otimes u_{\ell j} ]^{J_A} \rangle$.
Since Eq.~(\ref{eq3}) involves summation 
over all discrete Gamow states and integration over all 
scattering states along  the contour
$L_+^{\ell_j}$, the final result is 
independent of the s.p.  basis assumed. This is in contrast to standard SM
calculations where the model-dependence of SFs enters through the
specific choice of a s.p. state $a^+_{n \ell j}$.

%//////////////////////////////////////////////////////////////////////////////
\begin{figure}[hbt]
\begin{center}
\includegraphics[width=7.2cm]{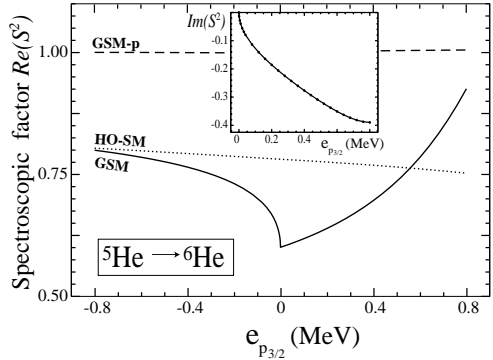}
\caption{The real part of the overlap
$\displaystyle \langle ^6{\rm He( g.s.}) | [^5{\rm He( g.s.}) 
\otimes p_{3/2}]^{0^+} \rangle^2$ in three different SM calculations as 
a function of the energy of the $0p_{3/2}$ resonant state.  
The solid line (GSM) shows  the full GSM result.
The dotted line (HO-SM) corresponds to the 
SM calculation in the oscillator basis of $0p_{3/2},0p_{1/2}$. 
The dashed line (GSM-p) shows the  GSM result in the pole approximation.
The imaginary part of $S^2$  is shown in the inset.
}
\label{fig_SF}
\end{center}
\end{figure}
%//////////////////////////////////////////////////////////////////////////////
In the following, we shall consider the ground state (g.s.) SF of  $^6$He
corresponding to  the channel
 $[{^5}{\rm He}({\rm g.s.})\otimes p_{3/2}]^{0^+}$,
where the single-neutron  $p_{3/2}$  channel  consists of the $0p_{3/2}$
resonant state and the $p_{3/2}$ non-resonant
scattering states on the complex contour $L_+^{p_{3/2}}$. The
s.p. basis is generated by a Woods-Saxon (WS) potential with
the ``$^{5}$He" parameter set \cite{Mic02} which reproduces the
experimental energies and widths of known  s.p. resonances $3/2_1^-$ and
$1/2_1^-$ in $^5$He. The GSM  Hamiltonian is a sum of the WS potential,
representing the inert $^4$He core, and the two-body
interaction among the valence neutrons. The latter is approximated
by a finite-range surface
Gaussian interaction \cite{Mic04} with  the range $\mu$=1\,fm
and the coupling constants depending on
the total angular momentum $J$ of the neutron pair: $V_0^{(0)}=-403$ MeV
fm$^{3}$, $V_0^{(2)}=-392$ MeV fm$^{3}$. These constants are fitted to
reproduce the g.s. binding energies of $^6$He and $^7$He in GSM. 

In order to investigate the continuum coupling, the depth of the WS
potential is varied so that the $0p_{3/2}$ s.p. state (the lowest
$p_{3/2}$ pole of the $S$-matrix), which is also the g.s. of $^5$He in
our model space, changes its character from bound to unbound. The
valence space for neutrons consists of the $0p_{3/2}$  resonant state,
complex-momentum  $p_{3/2}$ scattering continuum, and  non-resonant
$p_{1/2}$ scattering states along the real-$k$ axis. We do not consider
the $0p_{1/2}$ broad resonance  explicitly in the basis as it plays a
negligible role in the g.s. wave function of $^{6}$He \cite{Mic02}.  For
both $L_+^{\ell_j}$-contours, the maximal value for $k$ is
$k_{\rm{max}}=$3.27 fm$^{-1}$. The contours have been discretized with
up to 60 points and the attained precision on energies and widths is
better than 0.1 keV.

The calculated SF $[^5{\rm He}({\rm g.s.})\otimes p_{3/2}]^{0^+}$ in
$^6$He is shown in Fig. \ref{fig_SF} as a function of the energy of the
$0p_{3/2}$ pole. The SF strongly depends  on the position of the pole:
for $e_{0p_{3/2}}<0$ it decreases with $e_{0p_{3/2}}$ while  it 
increases  for $e_{0p_{3/2}}>0$. At the one-neutron (1n) emission
threshold in $^5$He, $e_{0p_{3/2}}$=0, the SF exhibits a cusp. At this
point, the derivative of SF becomes  discontinuous, and the coupling
matrix elements between the resonant $0p_{3/2}$ state  and the
non-resonant continuum reaches its maximum \cite{Op05}. Our calculations
are consistent  with the early estimate of the behavior of the cross
section near the threshold energy by Wigner \cite{wigner}. The
quickly varying component of SF {\it below} the 1n threshold  in 
$^5$He behaves as $(-e_{j\ell})^{\ell-1/2}$.  {\it Above} the 
threshold,  SF  is complex; the real part behaves as
$(e_{j\ell})^{\ell+1/2}$ while the imaginary part,
associated with the decaying nature of $^5$He,
  behaves as
$(e_{j\ell})^{\ell-1/2}$.

To assess the role of the  continuum, both resonant and non-resonant,
the GSM results are compared in
 Fig. \ref{fig_SF} with the traditional SM approach (HO-SM) and with the GSM in the pole
approximation (GSM-p). In the SM-HO variant, the harmonic oscillator (HO)
basis containing $0p_{3/2}$ and $0p_{1/2}$ states with $\hbar
\omega=41A^{-1/3}$ MeV was used. The s.p. energies were taken
from the GSM and their imaginary parts were ignored.
The resulting SF varies  little in the studied energy range 
and no threshold effect is seen. It is interesting to note
that in the limit of an appreciable binding, the $0p_{3/2}$ wave function is fairly well
localized, the importance of the continuum coupling is diminished, and 
the SM-HO result approaches the GSM limit.
In the GSM-p variant, only  the $0p_{3/2}$ and $0p_{1/2}$
resonant states are retained in the basis, i.e.,  the non-resonant scattering
states are ignored. Here, the g.s. of $^6$He is described by
 an almost  pure $[0p_{3/2} \otimes 0p_{3/2}]^{0^+}$ configuration
 and the resulting  SF is close to one in the whole energy region
considered. A dramatic difference between GSM and GSM-p results
illustrates the impact of the non-resonant continuum.

%//////////////////////////////////////////////////////////////////////////////
\begin{figure}[hbt]
\begin{center}
\includegraphics[width=7.2cm]{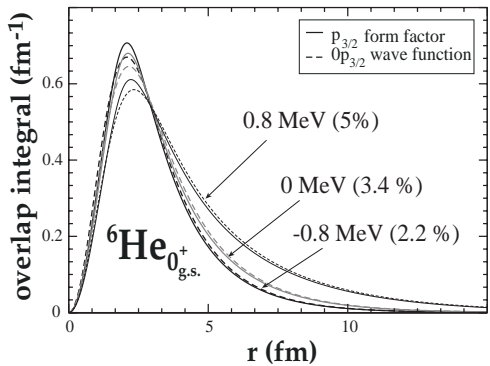}
\caption{Solid line: one neutron radial overlap integral 
for $\displaystyle \langle ^6{\rm He( g.s.}) | [^5{\rm He( g.s.}) 
\otimes p_{3/2}]^{0^+} \rangle^2$  in GSM
for three energies (--0.8, 0, 0.8 MeV) of the $0p_{3/2}$ resonant state.
Dashed line: radial wave function of the $0p_{3/2}$ resonant state of the WS
Hamiltonian adjusted to reproduce the 1n separation energy of $^6$He in GSM.
The relative difference between corresponding $\langle r^2 \rangle$ (in percents)
is also given.
}
\label{fig_FF}
\end{center}
\end{figure}
%//////////////////////////////////////////////////////////////////////////////
Figure~\ref{fig_FF} displays the associated  1n radial overlap integral calculated in the GSM
for three energies of the $0p_{3/2}$ resonant state. 
With $e_{0p_{3/2}}$ increasing,
the overlap becomes more diffused. For comparison, we also show the
radial wave function of the $0p_{3/2}$ resonant state of the WS
potential with a depth 
adjusted to reproduce the 1n separation energy of $^6$He 
calculated in GSM. The agreement between  s.p. wave functions and many-body 
overlap integrals is excellent.  Asymptotically, both quantities
fall down exponentially with a decay constant determined by
the 1n separation energy of $^6$He.
The effect of the non-resonant continuum 
is seen in a slightly better localization of GSM overlaps. 
(The  squared radii of overlap integrals are reduced by several percent
as compared to those of s.p. states.)

%//////////////////////////////////////////////////////////////////////////////
\begin{figure}[hbt]
\begin{center}
\includegraphics[width=7.2cm]{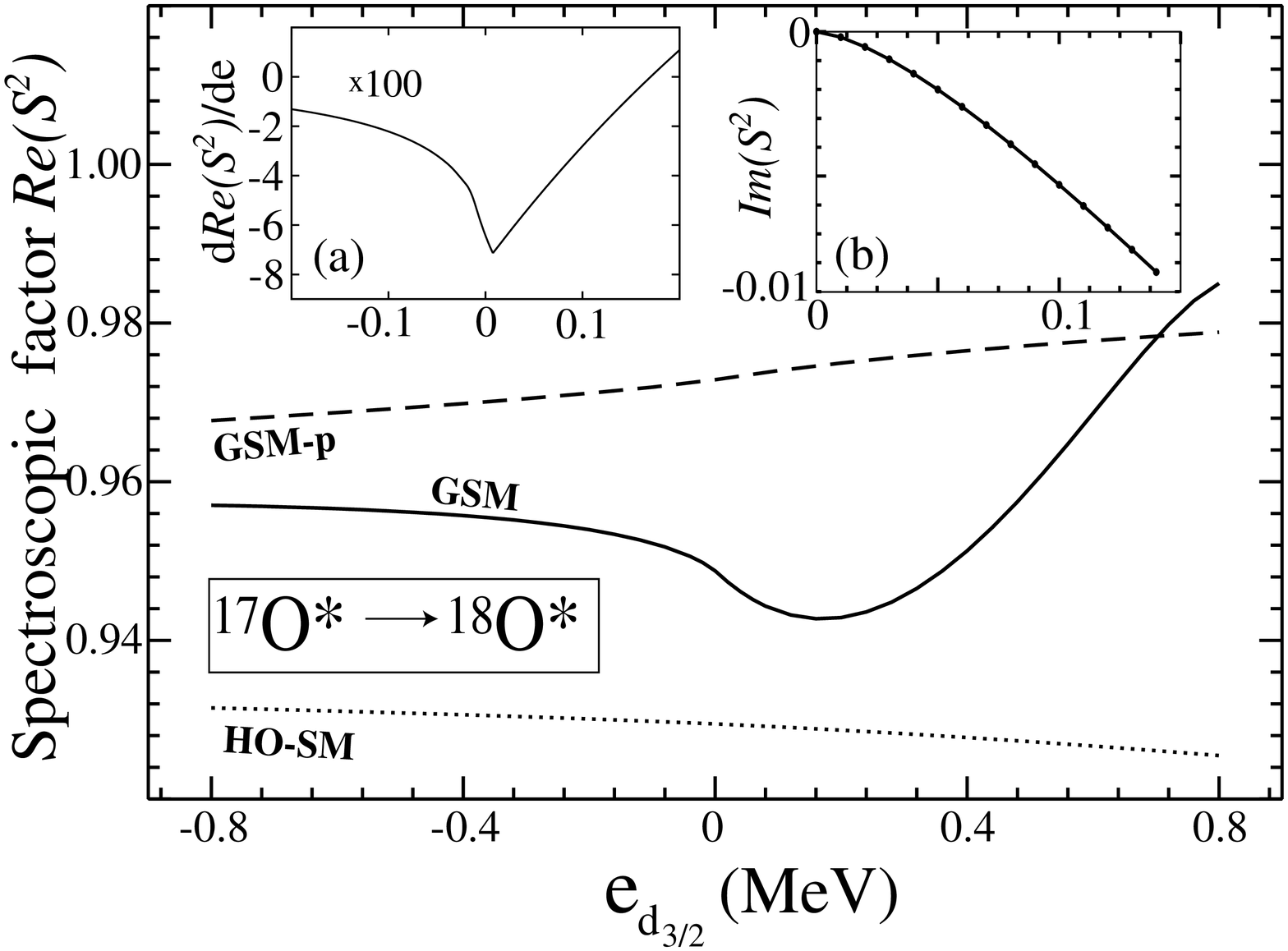}
\caption[]{Similar to Fig. \ref{fig_SF} except for the overlap 
 in the excited  $0^+_3$ state of $^{18}$O: $\displaystyle \langle
^{18}{\rm O}( 0^+_3) | [^{17}{\rm O}( 3/2_1^+) \otimes d_{3/2}]^{0^+}
\rangle^2$. The first derivative  of the SF in the neighborhood of the
$e_{0d_{3/2}}$=0 threshold  is shown
in the inset (a) while the inset (b) displays
the imaginary part of $S^2$.
}
\label{fig_SF1}
\end{center}
\end{figure}
%//////////////////////////////////////////////////////////////////////////////

To investigate the dependence of SFs on the orbital angular momentum
$\ell$ of the $S$-matrix pole, we show in Fig. \ref{fig_SF1} the SF for
the excited  $0^+_3$ state of $^{18}$O in the channel $[^{17}{\rm O}(
3/2_1^+) \otimes d_{3/2}]^{0^+}$. Here, the s.p. basis is generated by a
WS potential with the ``$^{17}$O" parameter set \cite{Mic02}  that
reproduces experimental energies and widths of the
$5/2_1^+$ and $1/2_1^+$ bound states and the $3/2_1^+$ resonance in
$^{17}$O. As a residual interaction, we took the surface delta
interaction \cite{Mic02} with the coupling constant $V_0$=--700 MeV
fm$^3$, which was fitted to reproduce the g.s. binding energy of $^{18}$O 
relative to the $^{16}$O core. The valence GSM space for neutrons
consists of the $0d_{5/2}$, $1s_{1/2}$,
and $d_{3/2}$ Gamow states  and the non-resonant complex $d_{3/2}$
continuum. The maximal value for $k$ on the contour $L_+^{d_{3/2}}$ is
$k_{\rm{max}}=$1.5 fm$^{-1}$. The contour has been discretized with 45
points and the resulting precision on energies and widths is $\sim0.1$ keV. 

The behavior of SF shown
 in Fig. \ref{fig_SF1} is similar to that of Fig.~\ref{fig_SF},
 except the variations are much smaller 
and the threshold behavior is different. Namely, the SF is continuous;
 it is its derivative that exhibits a cusp  around the $0d_{3/2}$ 
 threshold. Again, this is consistent with the general expectation 
that for $\ell$=2  $Re(S^2)$ {\em below} the 1n threshold of $^{17}$O
behaves as $(-e_{j\ell})^{3/2}$ while {\em above} the threshold
$Re(S^2)$ ($Im(S^2)$) should behave as $(e_{j\ell})^{3/2}$
($(e_{j\ell})^{5/2}$).
 The associated 1n $d_{3/2}$  radial overlap integrals (not displayed)
are extremely close to the s.p. resonant $0d_{3/2}$ wave functions;
the relative difference in square radii is around 1$\%$.

Coupling to the non-resonant scattering continuum may change
significantly energies of the many-body states  close to the
particle-emission threshold \cite{opr}.
We now shall examine the interplay between
the continuum coupling and
spin-orbit splitting  in $^5$He and $^7$He
using the same GSM Hamiltonian as before.
Since we are interested in the
first excited $1/2^-$ state of $^7$He, an  explicit inclusion of the 
$0p_{1/2}$ Gamow state is necessary.
 Consequently, the 
Berggren basis consists of 
the  $0p_{3/2}$ and $0p_{1/2}$ resonant
states and  their respective complex scattering contours $L_+^{p_{3/2}}$
and  $L_+^{p_{1/2}}$.  We take
$k_{\rm{max}}=$3.27 fm$^{-1}$ for both  contours, which 
were discretized with 27 points. The GSM eigenvalues for
$^7$He have been found using the Density Matrix Renormalization Group
technique \cite{dmrg1} adopted  for the non-hermitian GSM problem
\cite{dmrg2}. 

%//////////////////////////////////////////////////////////////////////////////
\begin{figure}[hbt]
\begin{center}
\includegraphics[width=7.2cm]{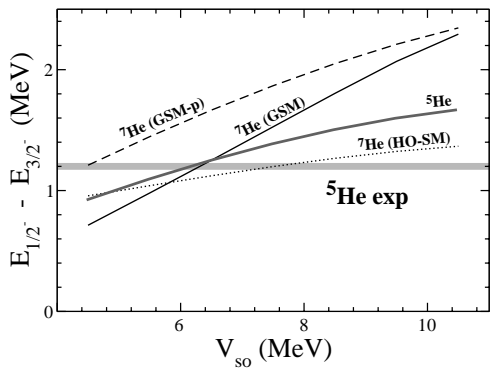}
\caption{The energy difference, $\Delta E$,  of  $1/2^-_1$ and $3/2^-_1$ states in
$^5$He and $^7$He as a function of the spin-orbit strength. 
The gray line  
shows $\Delta E$ in $^5$He. In this case $\Delta E$ is equivalent to
a s.p. spin-orbit splitting. The full GSM result for $^7$He is given by 
the black solid line, and 
dotted and dashed lines mark HO-SM and GSM-p results, respectively.
Experimental splitting  in $^5$He is indicated.}
\label{fig_SO}
\end{center}
\end{figure}
%//////////////////////////////////////////////////////////////////////////////
The energy difference $\Delta E = E_{1/2^-_1}-E_{3/2^-_1}$
of  the two lowest states in $^5$He and $^7$He,
calculated as a
function of the  spin-orbit term in the 
one-body potential, is shown in Fig.~\ref{fig_SO}.
In $^5$He, these two states in our model space are the s.p. Gamow
states $0p_{3/2}$ and $0p_{1/2}$, so that their energy difference is a
spin-orbit splitting.
In the absence of the residual
interaction, the same splitting is obviously obtained for $^7$He.

To investigate the importance of the non-resonant continuum,
we have performed the complete GSM calculation in the full
model space, as well as the GSM-p  and  HO-SM
calculations.  In
the pole approximation, $\Delta E$ follows closely the behavior for
$^5$He, albeit it is slightly larger. The calculated GSM-p states in
$^7$He  are almost the same as in
the absence of the residual interaction, so the 
$^5$He and $^7$He curves differ only by 
a small shift due to the residual
interaction. In the HO-SM variant,
the residual  interaction  gives rise to a 
configuration mixing, and $\Delta E$  in $^7$He is slightly reduced
as compared to the original spin-orbit splitting in $^5$He.
In the full GSM
calculations, the coupling to the non-resonant continuum comes to the
fore. The energy splitting in $^7$He varies  steeply with $V_{{\rm so}}$,
contrary to other curves. The difference between GSM and GSM-p results
reflects different correlation energies in the  $1/2^{-}_1$ and
$3/2^{-}_1$ states due to a coupling to the non-resonant continuum. This
means that effects of the one-body spin-orbit interaction are effectively
modified by the structure of loosely bound or unbound states, so
that the same interaction can lead to a
different  value of $\Delta E$ in
neighboring nuclei. 

In summary, our OQS calculations demonstrate the importance of
the non-resonant continuum for the spectroscopy of weakly bound nuclei.
Firstly, the GSM  study shows that
the behavior of SFs around the particle emission threshold 
exhibits a characteristic
non-perturbative behavior; the predicted  near-threshold reduction has, in our
model, nothing to do with short-range correlations. Secondly, the continuum
coupling strongly impacts the energetics of excited states (Thomas-Ehrman effect),
an effect that goes well beyond the traditional CQS description.
Interestingly, the effect of the non-resonant continuum on 
1n overlap integrals is minor: the single-particle approximation
often used in SM studies seems to work very well.

Useful discussions with the participants of the reaction part of 
the INT-05-3 Program
``Nuclear Structure Near the Limits of Stability" at the Institute
for Nuclear Theory, Seattle, are gratefully acknowledged. 
This work was supported by  the U.S. Department of Energy
under Contracts Nos. DE-FG02-96ER40963 (University of Tennessee),
DE-AC05-00OR22725 with UT-Battelle, LLC (Oak Ridge National
Laboratory), and DE-FG05-87ER40361 (Joint Institute for Heavy Ion
Research).

\end{document}